\documentclass[aps,reprint]{revtex4-1}
\usepackage{amsmath,amssymb}
\usepackage{amsfonts}
\usepackage{graphicx}
\usepackage{comment}
\usepackage{boxedminipage}
\usepackage{fancybox}
\usepackage{bm}
\usepackage{feynmf}

\begin{document}
\hfill{EPHOU-21-011, KEK-TH-2340}

\title{On Stringy Origin of Minimal Flavor Violation} 

\author{Tatsuo Kobayashi${}^{1}$}
\email[]{kobayashi@particle.sci.hokudai.ac.jp}
\author{Hajime Otsuka${}^{2}$}
\email[]{hotsuka@post.kek.jp}
\affiliation{${}^1$ Department of Physics, Hokkaido University, Sapporo 060-0810, Japan \\
${}^2$ KEK Theory Center, Institute of Particle and Nuclear Studies, KEK, 1-1 Oho, Tsukuba, Ibaraki 305-0801, Japan}


\begin{abstract}
We study the minimal flavor violation in the context of string effective field theory. Stringy selection rules indicate that $n$-point couplings among fermionic zero-modes and lightest scalar modes in the string effective action are given by a product of Yukawa couplings which are regarded as spurion fields of stringy and geometrical symmetries. 
Hence, Yukawa couplings determine the dynamics of flavor and CP violations. 
This observation strongly supports the hypothesis of minimal flavor violation in the Standard Model effective field theory. 
\end{abstract}

\maketitle

\section{Introduction}

Precision measurements of flavor and CP violations 
give us not only a precision test of the Standard Model (SM) but also 
clues of new physics beyond the SM. 
Current experimental searches put a strong constraint on the scale of new physics, 
namely the cutoff scale larger than the TeV scale in the context of low-energy 
effective field theory (EFT). 

The SMEFT approach is a powerful tool to capture signatures of new physics beyond the SM, 
and the flavor- and CP-violating processes are sensitive to higher-dimensional operators justified below the cutoff scale. 
Recalling the observation that the SM has an $U(3)^5$ flavor 
symmetry in the limit of vanishing Yukawa couplings, it is natural to assume that 
the SM Yukawa couplings are the only source of flavor and CP violations. 
In this hypothesis called Minimal Flavor Violation (MFV), 
Yukawa couplings are considered spurion fields under, e.g., the $U(3)^5$ maximal flavor symmetry \cite{Chivukula:1987py,DAmbrosio:2002vsn}; the higher-order couplings in the SMEFT are controlled by a product of Yukawa couplings to be invariant under the SM flavor and gauge symmetries. 
From the viewpoint of high-energy dynamics, the theory beyond the SM has to possess the non-trivial structure such that higher-dimensional operators generated by integrating out heavy states around the cutoff scale satisfy the criterion of the MFV hypothesis.

In the higher-dimensional theory, such as the string theory, only a limited class of $n$-point couplings is 
allowed in the four-dimensional (4D) EFT due to the existence of higher-dimensional gauge and Lorentz 
symmetries, and stringy selection rules. 
Explicit forms of $n$-point couplings were calculated in the string EFT on 
toroidal backgrounds, complex projective spaces, and some classes of Calabi-Yau manifolds with background gauge fields by utilizing conformal field theory (CFT) techniques and field theoretical methods. 
It turned out that higher-order couplings of matter zero-modes were constructed 
by the product of Yukawa couplings. 
For earlier attempts in the calculation of higher-order couplings, see Ref. \cite{Cvetic:2003ch} for Type IIA intersecting D-brane models, Ref. \cite{Cremades:2004wa} for Type IIB magnetized D-brane models, Ref. \cite{Hamidi:1986vh} for heterotic orbifold models, and Ref. \cite{Bershadsky:1993cx} for ${\cal N}=2$ topological string theory. 
Furthermore, the zero-mode solutions constitute a complete orthonormal system on general curved manifolds with background gauge fields; the $n$-point couplings of fermionic zero-modes and/or lightest scalar modes are determined by their $(n-1)$-point couplings in a similar way \cite{Honda:2018sjy}.

The phenomenological aspects of these higher-order couplings in the string EFT have not been fully investigated. 
The purpose of this paper is to explore the MFV hypothesis from the ultraviolet point of view by utilizing the structure of $n$-point couplings calculated in the string EFT. 
We find that various classes of string EFTs satisfy the criterion of MFV hypothesis, and flavor- and CP-violating processes are controlled 
by stringy selection rules.

\section{Higher-order couplings in string EFT}

We investigate the basic structure of $n$-point couplings in the string EFT framework 
from the MFV point of view.

\subsection{$n$-point couplings}

Let us begin with a class of 4D ${\cal{N}}=1$ EFTs arising from 
superstring theory compactified on toroidal as well as Calabi-Yau backgrounds.

On 4D ${\cal{N}}=1$ supersymmetric backgrounds, it has been known that the 
4D ${\cal{N}}=1$ superpotential of matter zero-modes are expressed by
\begin{align}
\begin{split}
W = y_{i_1 i_2 i_3}\Phi_{i_1} \Phi_{i_2} \Phi_{i_3} + \sum_{n=1} \frac{y_{i_1i_2\cdots i_{n+3}}}{\Lambda^n}\Phi_{i_1}\cdots \Phi_{i_{n+3}},
\label{eq:W}
\end{split}
\end{align}
where $\Phi_{i_n}$ denote matter chiral superfields originating from ten- and 
higher-dimensional gauge boson/gauginos for heterotic string theory and Type II string theory 
with D-branes, respectively. Here, $\Lambda$ represents the cutoff scale depending on the compactification scale. 

The higher-order couplings are explicitly calculated in toroidal orbifolds in the CFT as well as the 4D EFT. 
It is remarkable that $n$-point couplings of matter zero-modes 
take the form
\begin{align}
y_{i_1i_2\cdots i_{n}} = \sum_m y_{i_1i_2\cdots i_{n-2}m} \cdot y_{\bar{m}i_{n-1}i_{n}},
\label{eq:formula}
\end{align} 
for intersecting D6-brane models \cite{Cvetic:2003ch}, the T-dual magnetized D7-brane models \cite{Cremades:2004wa}, and heterotic orbifolds \cite{Hamidi:1986vh}. 
Here, we sum over the gauge-invariant modes $\Phi_m$ which do not necessarily correspond to zero-modes but include heavy states. 
This expression is inherited from operator product expansions in the two-dimensional CFT:
\begin{align}
    V_i (z) V_j (0) \sim \sum_k \frac{y_{ijk}}{z^{h_k -h_i -h_j}}V_k(0),
\end{align}
where $h_i$ denote the conformal dimensions of vertex operators $V_i$, corresponding to massless modes. 
Here, we rely on the perturbative calculation for $n$-point couplings, but non-perturbative 
effects such as D-brane instanton effects in Type IIA string theory lead to a similar structure of $n$-point couplings 
\cite{Blumenhagen:2006xt} protected by non-anomalous symmetries \cite{Hoshiya:2021nux}, where the coefficient includes the instanton action factor  $e^{-S_{\rm instanton}}$.

The explicit forms of $n$-point couplings are known on toroidal backgrounds, 
but it is challenging to derive $n$-point couplings on more general backgrounds 
such as Calabi-Yau threefolds. 
One of the difficulties in the calculation is due to the lack of analytical metric 
of entire Calabi-Yau threefolds and their submanifolds. 
However, in the heterotic string theory with standard embedding, $n$-point couplings 
of matter zero-modes can be extracted from the corresponding moduli couplings due to the 
identification between $SU(3)$ gauge bundle of heterotic $E_8$ gauge group and the tangent bundle 
of Calabi-Yau threefolds. It results in the same formula of $n$-point couplings (\ref{eq:formula}), 
where the $n$-point moduli couplings reduce to the derivative of Yukawa couplings with respect to 
the moduli fields \cite{Bershadsky:1993cx}. 
Furthermore, the equations of motion for matter zero-modes on general curved manifolds with background gauge fields also 
have a structure that $n$-point couplings among fermionic zero-modes and lightest scalar modes 
obey the form (\ref{eq:formula}) by utilizing the complete orthonormal system of Dirac and 
Klein-Gordon operators. 
The orthogonality of matter zero-modes $\Phi_i$ indicates that the lightest eigenstates of operators 
$D_i = \nabla_i -i A_i$ with background gauge fields $A_i$ satisfy the relation:
\begin{align}
\Phi_i \cdot \Phi_j \sim \sum_k y_{ijk}\cdot \Phi_k,
\end{align}
which results in an underlying structure of the formula (\ref{eq:formula}) in a way similar to the operator product 
expansions in the CFT. 

Note that the selection rule of $n$-point couplings (\ref{eq:formula}) will be applicable to the non-supersymmetric (SUSY) background. 
It means that the non-holomorphic $n$-point couplings of massless modes also obey the formula (\ref{eq:formula}) as 
confirmed in e.g., Ref. \cite{Honda:2018sjy}. 
For instance, four-fermion operators $y_{\bar{i}j\bar{k}l}\bar{\psi}_i\psi_j\bar{\psi}_k\bar{\psi}_l$ are generated by 
integrating out heavy states such as string modes as well as superpartners of SM fields. 

In this way, the selection rule of $n$-point couplings (\ref{eq:formula}) was explicitly demonstrated on toroidal and some classes of Calabi-Yau threefolds with background gauge fields and was also supported at the level of equations of motion \cite{Honda:2018sjy}. The structure of the formula (\ref{eq:formula}) will be understood from the viewpoint of string scattering amplitudes. 
For the tree-level string amplitude, the vertex operators inserted at the position of $i_1,...,i_{n-2}$ on Riemann surfaces correspond to massless modes, which join to give a massless or massive mode ($\Phi_m$) propagating in the Riemann surface and emit two massless modes $\Phi_{i_{n-1}}$ and $\Phi_{i_n}$. 
These observations strongly suggest that the decomposition of $n$-point couplings into Yukawa couplings holds on general grounds if they admit the CFT description.

\subsection{Minimal flavor violation}

We will consider the phenomenological interpretation of Eq. (\ref{eq:formula}) predicted 
in the string EFT. More concrete estimations of $n$-point couplings in the low-energy EFT will be presented in the next section. 

In the MFV hypothesis, Yukawa couplings are supposed to be spurion fields under 
the SM flavor symmetries, e.g., $U(3)^5$, and the SMEFT is constructed from the SM fields and Yukawa couplings, keeping the flavor and/or CP symmetries of quarks and leptons. The flavor-violating processes are then determined by the structure of Yukawa couplings. 
Remarkably, this treatment is realized in a certain class of string EFTs, since all the $n$-point couplings are decomposed into the product of Yukawa couplings, thereby yielding the magnitude of all the flavor- and CP-violating processes. 
Furthermore, $n$-point couplings are constrained by stringy selection rules and geometrical symmetries of the internal manifold in addition to the SM gauge symmetries, which allows us to restrict the unknown flavor symmetry of Yukawa couplings in the SMEFT prescription. 

In typical string models, matter zero-modes have representations under continuous and discrete symmetries associated with 
stringy selection rules and geometrical symmetries of the internal manifold, and the corresponding Yukawa couplings also have some representations under them.
In the string EFT, Yukawa couplings, as well as other couplings, are functions of moduli.
That corresponds to the important aspect of MFV; Yukawa couplings are spurion fields.
Since these moduli transform non-trivially under stringy and geometrical symmetries, 
couplings in the EFT and matter modes transform non-trivially.

One example is the modular symmetry $SL(2,\mathbb{Z})$ on toroidal (orbifold) backgrounds. 
Matter fields and their couplings transform non-trivially under the modular symmetry, but 
their Lagrangian is invariant. 
In particular, zero-mode wavefunctions of matter fields are described by the Jacobi theta function 
with modular weight 1/2 for D-brane models and 1 for heterotic orbifold models. 
Since $n$-point couplings are determined by the overlap integrals of $n$ number of matter zero-modes, 
these couplings are also described by modular forms as a function of the complex structure of the torus, which results in, e.g., $A^4$ or $S^4$ non-Abelian discrete flavor symmetries. 
(For more details, see Ref.~\cite{Ferrara:1989qb} for heterotic orbifold models, and Ref.~\cite{Kobayashi:2018rad} for magnetized D-brane models.)
In multi-moduli theories, the modular symmetry is enlarged to the symplectic modular symmetry $Sp(2h,\mathbb{Z})$, 
which governs the flavor and CP structures. 
For instance, in the heterotic string theory with standard embedding on Calabi-Yau threefolds, 
Yukawa and higher-order couplings computed by taking an appropriate number of derivatives of the prepotential are tensor representations of a subgroup of $Sp(2h,\mathbb{Z})$ in the Calabi-Yau moduli space (such as $S_4$ flavor symmetry \cite{Ishiguro:2021ccl}).  
These couplings depend on the complex structure and K\"ahler moduli for the fundamental and anti-fundamental representations of $E_6$ gauge group, respectively. 

Another example is given by heterotic orbifold models, 
where stringy selection rules and geometrical symmetries lead to 
flavor symmetries such as $D_4$ and $\Delta(54)$ \cite{Kobayashi:2004ya}.
Twisted modes are non-trivial representations of these symmetries, while 
untwisted modes are trivial singlet. 
Yukawa couplings, as well as other couplings, are also non-trivial representations 
when twisted modes develop their vacuum expectation values corresponding to the 
blow-up of orbifold singularities.

Starting from the supersymmetric constructions, the SUSY-breaking terms will cause the flavor violating processes in the low-energy EFT. 
In the typical string models, the rank of Yukawa couplings is 1 in the perturbative regime, i.e., the large volume and weak string coupling regime \cite{Ibanez:2012zz}. 
It corresponds to the $U(2)$ flavor symmetry acting on first-two light generations. 
In this way, the flavor violations induced by the soft SUSY-breaking terms are approximately suppressed. Also, we can study more generic models, where Yukawa matrices correspond to rank-3 matrices. 
The soft scalar masses $m_{\Phi_{i}}^2$ and the $A$-terms induced by the SUSY-breaking fields $X$ are written by \cite{Kaplunovsky:1993rd},
\begin{align}
    m_{\Phi_{i}}^2 &= m_{3/2}^2 - \sum_X |F^X|^2 \partial_X\partial_{\bar{X}}\ln \partial_{\Phi_{i}} \partial_{\bar{\Phi}_i}K_{\rm matter},
    \nonumber\\
    A_{ijk} &= A_{i} + A_{j} +A_{k} - \sum_X \frac{F^X}{y_{ijk}} \partial_X (y_{ijk}),
\end{align}
with
\begin{align}
    A_i = \sum_X F^X \partial_X \ln \left(e^{-K_{\rm mod}} \partial_{\Phi_{i}} \partial_{\bar{\Phi}_i}K_{\rm matter}\right),
\end{align}
where $m_{3/2}$ is the gravitino mass, $K_{\rm mod}$ and $K_{\rm matter}$ denote the moduli and matter K\"ahler potentials, respectively. 
Here, we assume that the matter K\"ahler metric is a diagonal form. 
When $X$ are moduli fields, soft terms are also invariant under geometrical symmetries of the internal manifold, such as the modular symmetries, since moduli fields and their $F$-terms are the same representations under the modular symmetries.

In this way, various classes of string EFTs satisfy the requirement of MFV hypothesis, and the flavor structure of higher-order couplings is determined by Yukawa couplings reflecting geometrical symmetries of the internal manifold. 
The representations of higher-order couplings under the geometrical symmetries are given by the tensor product of Yukawa couplings in a way similar to the MFV. Furthermore, the magnitudes of CP violation in Yukawa and higher-order couplings have a common origin in string axions associated with the complex structure or K\"ahler moduli of toroidal and Calabi-Yau geometries.

Even if string EFTs satisfy the MFV at the compactification scale, 
various physical stages may occur between the compactification scale and low energy one, (i) some modes become massive, and (ii) some scalar fields develop their vacuum expectation values.
At the stage (i), we integrate out massive modes $\Phi_m$, and 
then new operators appear after the integration keeping the rule (\ref{eq:formula}).
At the stage (ii), new couplings are induced by vacuum expectation values of scalar fields. 
For example, suppose that there is the coupling, $y_{ijk\ell}\Phi_i \Phi_j \Phi_k \Phi_\ell$ and $\Phi_i$ develops its vacuum expectation value. 
Then, a new coupling appears as $y'_{jk\ell} \Phi_j \Phi_k \Phi_\ell$ with $y'_{jk \ell}=y_{ijk\ell}\langle \Phi_i \rangle$. 
Since the coupling constant $y_{ijk\ell}$ is the spurion depending on moduli, the new coupling constant $y'_{jk \ell}$ is also the spurion as functions of $\langle \Phi_i \rangle$ as well as moduli.
The transformation behavior of $y'_{jk \ell}$ under (flavor and CP) symmetries is inherited from one of $y_{ijk\ell}\langle \Phi_i \rangle$.
Thus, the MFV framework is not violated and remains in the low-energy EFT, although the cutoff scale may shift from the compactification scale to the scale including $\langle \Phi_i \rangle$ as well as mass scale of massive modes.
This statement would hold true for some fermion condensations.
Thus, the MFV framework can be realized in various classes of string EFTs, 
and remain in the low-energy EFT.

\section{The structure of higher-order couplings in the low-energy EFT}

We discuss the phenomenological implications of the formula (\ref{eq:formula}) 
into higher-dimensional operators relevant to flavor physics in the low-energy EFT.

\subsection{Dimension-5 operators}

Here, we begin with a dimension-5 operator of the 4D ${\cal{N}}=1$ effective action realizing 
the Minimal Supersymmetric Standard Model (MSSM) at the low-energy scale. 
The dimension-5 operator invariant under $SU(2)_L\times U(1)_Y$ is the Weinberg operator:
\begin{align}
 \frac{y_{L_i HL_j H}}{\Lambda}(L_i H)(L_j H).
\end{align}
The formula (\ref{eq:formula}) indicates that the four-point couplings $y_{L_i HL_j H}$ are determined by the Yukawa couplings including the lepton sector: $y_{L_i HL_j H} \simeq \sum_m y_{L_i \Phi_m H}\times y_{L_j \bar{\Phi}_m H}$, 
independently on whether right-handed neutrinos appear below
the compactification scale or not.

The SM or its supersymmetric extension can be engineered using several stacks of D-branes, e.g., intersecting D-brane models. 
In the context of heterotic string theory, the model building of grand unified theories (GUTs) has been based on $SU(5)$, $SO(10)$, and $E_6$ GUTs at the intermediate scale, which is subsequently broken to the SM gauge group by a Wilson line. 
The selection rules of $n$-point couplings constrain the higher-dimensional operators not only in the SM 
but also in GUTs. 
For instance, the dimension-5 operators for fundamental representations in the SUSY $E_6$ GUT 
\begin{align}
\frac{a_{ijkl}}{\Lambda}27_i 27_j 27_k 27_l
\end{align}
are given by $a_{ijkl}=\partial_i \partial_j \partial_k \partial_l {\cal F}$ with ${\cal F}$ being the prepotential of 
Calabi-Yau threefolds, where $a_{ijkl}$ are generated by the contribution of massive modes in the formula (\ref{eq:formula}). 
In the large complex structure regime, the prepotential takes the form:
\begin{align}
{\cal F} = {\cal F}_{\rm poly} +{\cal F}_{\rm inst},
\end{align}
where ${\cal F}_{\rm poly}$ denotes the cubic polynomial functions with respect to the complex 
structure moduli, and ${\cal F}_{\rm inst}$ represents quantum corrections corresponding to worldsheet instanton effects in the topological A-model. 
Since higher-dimensional operators have an origin in the quantum corrections, 
these exponentially suppressed corrections will not be significant for the flavor- and CP-violating processes. 

The couplings of dimension-5 proton decay operators in the SUSY $SU(5)$ GUT
\begin{align}
    \frac{b_{ijk\bar{l}}}{\Lambda}10_i 10_j 10_k \bar{5}_l
\end{align}
are also written by the product of Yukawa couplings:
\begin{align}
    b_{ijk\bar{l}} = \sum_m y_{10_i10_j \Phi_m} \times y_{10_k \bar{5}_l \bar{\Phi}_m},
\end{align}
where $\Phi_m$ will include GUT Higgs fields or other heavy states. 
If these operators are prohibited by the existence of hidden $U(1)$ symmetries, 
$SU(5)$ singlet fields appear in the proton decay operators in a gauge-invariant way. 
Even in this case, the couplings of proton decay operators would be restricted to satisfy 
the relation (\ref{eq:formula}). 
On the toroidal backgrounds, the four-point couplings are given by the Jacobi theta function with respect to the complex structure of the torus $\tau$. Since such an elliptic function behaves as 
$e^{2\pi i \tau}$, these higher-order couplings can also be suppressed in the large complex structure regime ${\rm Im}(\tau)\gg 1$.

\subsection{Dimension-6 flavor-violating operators}

Let us study the dimension-6 flavor-violating operators in the SMEFT following Ref. \cite{Grzadkowski:2010es}, including the 1350 CP-even and 1149 CP-odd operators in the absence of any flavor symmetry \cite{Alonso:2013hga}. 
Concerning the recent flavor anomalies such as $B$ physics, we focus on four-fermion operators: the $\Delta F=2$ quark operators
\begin{align}
    \frac{c_{\bar{i}j\bar{k}l}}{\Lambda^2}(\bar{Q}_i \gamma_\mu P_{L,R} Q_j)(\bar{Q}_k \gamma^\mu P_{L,R} Q_l),
\end{align}
and the $\Delta F=1$ semileptonic operators 
\begin{align}
    &\frac{d_{\bar{i}j\bar{k}l}}{\Lambda^2}(\bar{Q}_i \gamma_\mu P_{L,R} Q_j)(\bar{L}_k \gamma^\mu P_{L,R} L_l),
\end{align}
with the 4D Gamma matrix $\gamma_\mu$, and the projective operators $P_{L,R}$, 
where we pick up specific four-fermion operators for illustrative purposes. 
Given the superpotential (\ref{eq:W}), the four-fermion operators are generated by the spontaneous SUSY breaking or the contribution of stringy states, and couplings $\{c_{\bar{i}j\bar{k}l}, d_{\bar{i}j\bar{k}l}\}$ are decomposed into 
the product of moduli-dependent three-point couplings: $\sum_m y_{\bar{i}jm}y_{\bar{m}\bar{k}l}$ where the virtual mode $\phi_m$ corresponds 
to heavy superpartners or stringy modes appearing in the gauge-invariant way. (For the derivation of dimension-6 operators in the case of matching the MSSM onto the SMEFT, e.g., Ref. \cite{Wells:2017vla}.)  
Here, the cutoff scale determined by the SUSY-breaking scale or the string scale should be larger than at least a few 
TeV scale due to recent flavor experiments, e.g., various meson mixing in $\Delta F=2$ processes \cite{Aebischer:2020dsw} and the lepton-flavor universality violation in the semileptonic decay processes of the $B$-meson \cite{Celis:2017doq}. (For more details about the constraints on dimension-6 SMEFT operators, see, e.g., Ref. \cite{Isidori:2010kg} and references therein.) 
Furthermore, the rank 1 Yukawa couplings are typically realized at the leading order of string EFTs \cite{Ibanez:2012zz}, corresponding to the $U(2)$ flavor symmetry acting on first-two light generations. This setup provides the useful EFT approach for the recent flavor anomalies \cite{Barbieri:2011ci}.

When the four-fermion operators are generated by heavy superpartner contributions, the Yukawa couplings of quarks and leptons are constrained by a realization of their mass hierarchies and suppression of dangerous flavor and CP violations. 
Such a hierarchical structure would be possible at a specific point of moduli space such as the large complex structure regime, and the vacuum expectation values of moduli and axion fields appearing in Yukawa couplings will be fixed by the stabilization mechanism of the internal manifold such as flux compactifications with or without non-perturbative effects developed in Type IIB string theory \cite{Giddings:2001yu,Kachru:2003aw,Balasubramanian:2005zx}.


\subsection{Derivative couplings}

In addition to the SM flavor and gauge symmetries, the gauge and higher-dimensional Lorentz symmetries constrain the form of $n$-point couplings with derivatives. The derivative couplings in the higher-dimensional theory such as $\psi \bar{\psi}|D_M\phi|^2$ lead to the 4D couplings through the compactification: $\chi \bar{\chi}|D_\mu \xi|^2$, where $\chi$ and $\xi$ are 4D fermions and scalars with the 4D coordinates $x^\mu$, originating from higher-dimensional fermions $\psi$ and scalars $\phi$ with the coordinate in higher-dimensional spacetime $x^M$, respectively. 
Similarly, the coefficient of $\chi \bar{\chi}|D_\mu \xi|^2$ is the same form of $\chi \bar{\chi}|\xi|^2$ thorough the 
compactification of the internal space. 
When we identify $\xi$ with the Higgs fields, the coupling of $\chi \bar{\chi}|D_\mu \xi|^2$ operator corresponds to the four-point coupling of fermionic zero-modes and Higgs fields. 
In this way, the derivative couplings are also governed by $n$-point couplings and satisfy the criterion of the MFV.  

Note that these analyses would be applicable to other derivative couplings with or without multiple scalar fields such as multi-Higgs models, which are often predicted in the string effective action.  
The effective action, as well as the flavor structure of the multiple Higgs fields, will also be constrained by the formula (\ref{eq:formula}) (see, e.g., Ref. \cite{Branco:1996bq}).


\section{Conclusions}

It has been known that the MFV hypothesis is a powerful tool to 
capture signatures of flavor and CP violations, but 
its high-energy behavior has not been established. 
In this paper, we focused on the properties of higher-order couplings 
in a certain class of string EFTs. 
We have shown that the stringy structure naturally satisfies the 
criterion of MFV hypothesis; all the flavor- and CP-violating processes 
are controlled by the product of Yukawa couplings. 
Furthermore, moduli fields appearing in Yukawa couplings transform non-trivially 
under stringy and geometrical symmetries, implying that Yukawa couplings are regarded as spurion fields under them. 
The MFV scenario also works in the low-energy EFT below the compactification scale.

In the context of string EFT, the appearance of flavor symmetries depends 
on the structure of Yukawa couplings obtained through the compactification of extra-dimensional spaces. 
One of the interesting symmetries is the modular symmetry.
For the toroidal backgrounds, the modular symmetry $SL(2,\mathbb{Z})$ 
and its finite subgroups such as $A_4$ and $S_4$ play an 
important role in the flavor structure of quarks and leptons, where the 
Yukawa couplings have non-trivial representations under the modular 
symmetry. A similar phenomenon appears in the heterotic string theory with standard embedding on Calabi-Yau threefolds. 
Other interesting symmetries can be provided by stringy selection rules and geometrical symmetries of 
compact spaces.
It would be fascinating to examine in detail the flavor and CP violations induced by higher-order 
couplings such as the semileptonic decays of $B$-mesons on concrete models and exemplify the realization of MFV hypothesis on other corners of 
string models.  

Furthermore, we have only focused on the classical $n$-point couplings of matter fields. 
It would be interesting to incorporate quantum corrections into the matter couplings 
and check the validity of the formula (\ref{eq:formula}) against these corrections.


\begin{acknowledgments}
  We would like to thank Y. Honma, M. Tanimoto and K. Yamamoto for useful discussions and comments.
  H. O. was supported in part by JSPS KAKENHI Grant Numbers JP19J00664 and JP20K14477.
\end{acknowledgments}


\end{document}